\documentclass[aps,prb,longbibliography,reprint,superscriptaddress]{revtex4-2}
\usepackage{siunitx, upgreek}
\usepackage{graphicx}
\usepackage[english]{babel}
\usepackage{hyperref}
\usepackage[capitalize]{cleveref}
\usepackage{float}
\usepackage{xcolor}
\hypersetup{
    colorlinks,
    linkcolor={blue!50!black},
    citecolor={blue!50!black},
    urlcolor={blue!80!black}
}
\begin{document} 
%
\title{THz-Driven Coherent Magnetization Dynamics in a Labyrinth Domain State} 
%
\author{Matthias~Riepp}
\email[E-mail: ]{matthias.riepp@ipcms.unistra.fr}
\affiliation{Institut de Physique et Chimie des Matériaux de Strasbourg, UMR 7504, Université de Strasbourg, CNRS, 67000 Strasbourg, France}
\affiliation{Deutsches Elektronen-Synchrotron~DESY, Notkestra{\ss}e~85, 22607~Hamburg, Germany}
\author{André~Philippi-Kobs}
\affiliation{Deutsches Elektronen-Synchrotron~DESY, Notkestra{\ss}e~85, 22607~Hamburg, Germany}
\affiliation{Institut für Experimentelle und Angewandte Physik, Christian-Albrechts-Universit{\"a}t zu Kiel, Leibnitzstraße 19, 24098 Kiel, Germany}%
\author{Leonard~M{\"u}ller}
\affiliation{Deutsches Elektronen-Synchrotron~DESY, Notkestra{\ss}e~85, 22607~Hamburg, Germany}
\author{Wojciech~Roseker}
\affiliation{Deutsches Elektronen-Synchrotron~DESY, Notkestra{\ss}e~85, 22607~Hamburg, Germany}
\author{Rustam~Rysov}
\affiliation{Deutsches Elektronen-Synchrotron~DESY, Notkestra{\ss}e~85, 22607~Hamburg, Germany}
\author{Robert~Fr{\"o}mter} 
\affiliation{Institute of Physics, Johannes Gutenberg-University Mainz, 55099 Mainz, Germany}
\author{Kai~Bagschik}
\affiliation{Deutsches Elektronen-Synchrotron~DESY, Notkestra{\ss}e~85, 22607~Hamburg, Germany}
\author{Marcel~Hennes}
\affiliation{Institut des NanoSciences de Paris, UMR7588 , Sorbonne Université, CNRS, 75005~Paris, France} 
\author{Deeksha~Gupta}
\affiliation{Institut de Physique et Chimie des Matériaux de Strasbourg, UMR 7504, Université de Strasbourg, CNRS, 67000 Strasbourg, France}
\author{Simon~Marotzke}
\affiliation{Deutsches Elektronen-Synchrotron~DESY, Notkestra{\ss}e~85, 22607~Hamburg, Germany}
\affiliation{Institut für Experimentelle und Angewandte Physik, Christian-Albrechts-Universit{\"a}t zu Kiel, Leibnitzstraße 19, 24098 Kiel, Germany}
\author{Michael~Walther}
\affiliation{Deutsches Elektronen-Synchrotron~DESY, Notkestra{\ss}e~85, 22607~Hamburg, Germany}
\author{Sa\v{s}a~Bajt} 
\affiliation{Center for Free-Electron Laser Science CFEL, Deutsches Elektronen-Synchrotron DESY, 22607 Hamburg, Germany}
\affiliation{The Hamburg Centre for Ultrafast Imaging, 22761 Hamburg, Germany}
\author{Rui~Pan}
\affiliation{Deutsches Elektronen-Synchrotron~DESY, Notkestra{\ss}e~85, 22607~Hamburg, Germany}
\author{Torsten~Golz}
\affiliation{Deutsches Elektronen-Synchrotron~DESY, Notkestra{\ss}e~85, 22607~Hamburg, Germany}
\author{Nikola~Stojanovic} 
\affiliation{Institute for Optical Sensor Systems, Deutsches Zentrum f{\"u}r Luft- und Raumfahrt, Rutherfordstraße 2, 12489 Berlin, Germany}
\author{Christine~Boeglin}
\affiliation{Institut de Physique et Chimie des Matériaux de Strasbourg, UMR 7504, Université de Strasbourg, CNRS, 67000 Strasbourg, France}
\author{Gerhard~Gr{\"u}bel}
\altaffiliation[Present address: ]{European X-Ray Free-Electron Laser Facility GmbH, Holzkoppel 4, 22869 Schenefeld, Germany}
\affiliation{Deutsches Elektronen-Synchrotron~DESY, Notkestra{\ss}e~85, 22607~Hamburg, Germany}
%
%
\begin{abstract}
Terahertz~(THz) light pulses can be used for an ultrafast coherent manipulation of the magnetization. Driving the magnetization at THz~frequencies is currently the fastest way of writing magnetic information in ferromagnets. Using time-resolved resonant magnetic scattering, we gain new insights to the THz-driven coherent magnetization dynamics on nanometer length scales. We observe ultrafast demagnetization and coherent magnetization oscillations that are governed by a time-dependent damping. This damping is determined by the interplay of lattice heating and magnetic anisotropy reduction revealing an upper speed limit for THz-induced magnetization switching. We show that in the presence of nanometer-sized magnetic domains, the ultrafast magnetization oscillations are associated with a correlated beating of the domain walls. The overall domain structure thereby remains largely unaffected which highlights the applicability of THz-induced switching on the nanoscale.
\end{abstract}
\maketitle
%

%
\section{INTRODUCTION}
%
Understanding the magnetization dynamics driven by ultrashort light pulses is of key importance for developing faster and more energy efficient opto-magnetic memory technologies. A promising way to a controlled manipulation of the magnetization in ferromagnetic thin films on ultrafast time scales is the use of light pulses with frequencies in the terahertz~(THz) regime ($\nu_\mathrm{THz}\approx 0.1$--$10\cdot10^{12}\,\mathrm{Hz}$)~\cite{kampfrath_resonant_2013, walowski_perspective_2016, barman_2021_2021}. In contrast to incoherent ultrafast demagnetization induced by femtosecond optical laser pulses with frequencies in the infrared~(IR) regime ($\nu_\mathrm{IR}\approx 10^{14}\,\mathrm{Hz}$)~\cite{beaurepaire_ultrafast_1996}, the electric field component $E_\mathrm{THz}$ is capable of driving a coherent ultrafast demagnetization with significantly lower energy transfer to the sample~\cite{pellegrini_ab_2022}. Moreover, the magnetic field component $H_\mathrm{THz}$ may induce coherent oscillations~\cite{kampfrath_coherent_2011, bonetti_thz-driven_2016, shalaby_coherent_2018, neeraj_inertial_2021} and, at high intensities, even a switching of the magnetization~\cite{back_magnetization_1998, tudosa_ultimate_2004, polley_thz-driven_2018, neeraj_magnetization_2022}. The possibility of exciting coherent magnetization dynamics at THz~frequencies in ferromagnets, i.\,e., far from the ferromagnetic precession resonance, was explained by the inertia of the magnetization~\cite{ciornei_magnetization_2011, bhattacharjee_atomistic_2012, olive_deviation_2015, mondal_relativistic_2017-1}. As a consequence, the magnetization may undergo \emph{nutation} dynamics, i.\,e., a trembling of the magnetization vector at THz~frequencies respectively femtosecond time scales. Ultrashort THz~light pulses therefore promise high-speed and low-power-consumption information writing in ferromagnets. 

So far, experiments mainly addressed THz-driven ultrafast magnetization dynamics in homogeneously magnetized thin films, i.\,e., in the single-domain state~\cite{vicario_off-resonant_2013, bonetti_thz-driven_2016, shalaby_coherent_2018, hudl_nonlinear_2019, unikandanunni_inertial_2022, rouzegar_laser-induced_2022, unikandanunni_inertial_2022, neeraj_terahertz_2022, salikhov_coupling_2023}. Information on the THz-driven magnetization dynamics in non-uniformly magnetized states, such as the nanoscale multi-domain states in Co/Pt~multilayers with perpendicular magnetic anisotropy~(PMA), is still lacking. A dependence of the THz-driven coherent magnetization dynamics on the magnetic anisotropy energy~(MAE) was discovered recently by investigating Co~thin films with fcc, bcc and hcp crystal structure~\cite{unikandanunni_inertial_2022}.

%
\begin{figure*}
{\includegraphics[width=0.98\textwidth]{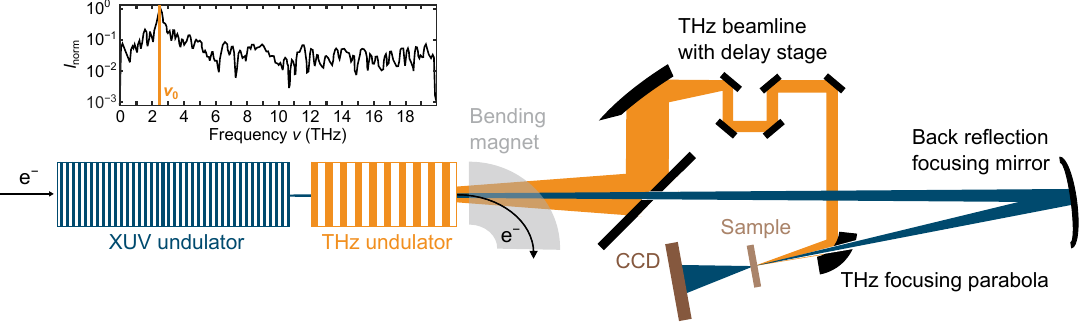}
\caption{\label{Fig1_Setup}
\textbf{Schematics of tr-XRMS at the BL3~instrument of FLASH \quad} Relativistic electron bunches consecutively traverse the XUV and THz~undulator producing intrinsically synchronized pump and probe pulses. In a custom-made end station, time-delayed THz-pump and XUV-probe pulses are focused quasi-collinearly onto the sample via a parabolic mirror and a back-reflection multilayer mirror, respectively. Included is the polychromatic THz~pump spectrum with a fundamental frequency $\nu_0 = \SI{2.5}{THz}$ measured by electro-optical sampling~(EOS).
}}
\end{figure*}
In this article, we present THz-driven coherent magnetization dynamics in the labyrinth domain state of a Co/Pt~multilayer with PMA. We employ time-resolved XUV resonant magnetic scattering~(tr-XRMS) at the free-electron laser~(FEL) FLASH to resolve these coherent dynamics with femtosecond time and nanometer spatial resolution~\cite{stohr_magnetism_2006, gutt_resonant_2009, gutt_single-pulse_2010, vodungbo_laser-induced_2012, wang_femtosecond_2012}. The magnetization shows different responses depending on the used THz~pump fluence. For low-fluence excitation with a filtered THz~spectrum ($\nu<\SI{6.0}{THz}$), the magnetization undergoes an ultrafast quenching and recovery within \SI{1}{ps}. For high-fluence excitation with the full THz~spectrum, a step-like quenching within \SI{2}{ps} occurs followed by oscillatory dynamics in resonance with the THz~fundamental frequency $\nu_0 = \SI{2.5}{THz}$. The data are consistent with incoherent and coherent ultrafast magnetization dynamics driven by the $E_\mathrm{THz}$- and $H_\mathrm{THz}$-field components. However, a time-dependent damping has to be introduced which is modeled by the interplay of lattice heating and PMA reduction. The oscillatory magnetization dynamics are associated with correlated dynamics of the domain state's form factor, interpreted as a successive broadening and narrowing of the domain walls, whereas the overall domain structure is conserved.

%
\section{RESULTS AND DISCUSSION}
%
\textbf{Experimental details} The THz-driven magnetization dynamics were studied by tr-XRMS at the BL3~instrument of FLASH~(see Methods). The schematics of the experiment are shown in Fig.\,\ref{Fig1_Setup}. The planar electro-magnetic THz~undulator at BL3 with nine full periods was tuned to generate pump pulses with a fundamental frequency $\nu_0=\SI{2.5}{THz}$ which results in a pump-pulse duration of~\SI{3.6}{ps}. Importantly, the pump pulses contain a broad frequency spectrum, in particular, also high-frequency components reaching up to the IR regime~(see Fig.\,\ref{Fig1_Setup}). We call this the unfiltered THz~radiation. From a pump-pulse intensity of~$\SI{23}{\upmu J}$ measured by a radiometer~\cite{pan_photon_2019} and a beam size of~$200\times160\,\mathrm{\upmu m^2}$ measured by a fluorescent screen at the sample position, the calculated pump fluence is $F_\mathrm{THz}=\SI{92}{mJ\,cm^{-2}}$. This corresponds to electric and magnetic field amplitudes $E_\mathrm{THz}=\SI{4}{MV\,cm^{-1}}$ and $\mu_0H_\mathrm{THz}=\SI{1.4}{T}$, respectively. Alternatively, a longpass filter that blocks frequency components $\nu \gtrsim \SI{6.0}{THz}$ was inserted in the THz~beamline. For this filtered THz~radiation, the pump~fluence is reduced at least by a factor of four~\cite{liu_investigating_2022}.

%
\begin{figure}
{\includegraphics[width=0.48\textwidth]{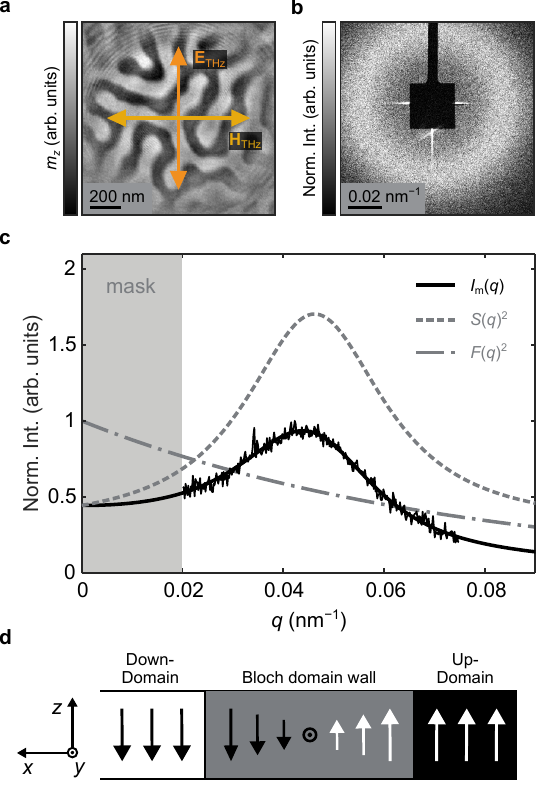}
\caption{\label{Fig2_ScattAnalysis}
\textbf{Processing tr-XRMS data \quad a}~Fourier-transform holography image of a typical labyrinth domain state $m_z(\mathbf{r})$ in Co/Pt~multilayers showing up- and down-magnetized domains as dark and bright contrast. Arrows indicate the propagation directions of the electric and magnetic field components $\mathbf{E}_\mathrm{THz}(t)=(0,\,E_y(t),\,0)$ and $\mathbf{H}_\mathrm{THz}(t)=(H_x(t),\,0,\,0)$. \textbf{b}~Normalized magnetic scattering image $I_\mathrm{m}(\mathbf{q}, t=\SI{-1}{ps})$ obtained by tr-XRMS from the labyrinth domain state of the [Co/Pt]$_8$~multilayer used in this experiment. \textbf{c}~Corresponding azimuthal average of the magnetic scattering intensity $I_\mathrm{m}(q)$. Included are a fit to the data using eq.\,(\ref{Eq_Lorentzian_Zusin}) and its individual contributions, i.\,e., the form factor $F(q)^2$ and the structure factor $S(q)^2$. The $I_\mathrm{m}(q)$ and $S(q)^2$ are normalized to the maximum of $I_\mathrm{m}(q)$ for clarity. \textbf{d}~1D illustration of the individual contributions to $I_\mathrm{m}(q)$ in real space: Bloch domain walls correspond to the magnetic unit cell, up- and~down-magnetized domains to the magnetic lattice.
}}
\end{figure}
A typical labyrinth domain state $m_z(\mathbf{r})$ in Co/Pt~multilayers with PMA is shown in Fig.\,\ref{Fig2_ScattAnalysis}\,\textbf{a}. Here, $m_z = M_z/M_\mathrm{s}$ is the $z$-component of the magnetization normalized to its value at saturation. Note that for THz~pump pulses incident normally on such an OOP domain state, the Zeeman torque $\mathbf{T}=\mathbf{m}\times\mathbf{H}$ is maximized. The scattered intensity $I(\mathbf{q}, t=\SI{-1}{ps})$, obtained by tr-XRMS from the labyrinth domain state of the [Co/Pt]$_8$~multilayer used in this experiment~(see Methods), is shown in Fig.\,\ref{Fig2_ScattAnalysis}\,\textbf{b}. In the kinematical limit using linearly polarized light incident normally on a thin film with PMA, the scattered intensity is given by pure charge and pure magnetic scattering contributions $I(\mathbf{q})=I_\mathrm{c}(\mathbf{q})+I_\mathrm{m}(\mathbf{q})$~\cite{kortright_resonant_2013}. Recently it was shown that the charge scattering contribution is orders of magnitude smaller as compared to the first-order magnetic scattering contribution in the here investigated region of $q$-space~\cite{zusin_ultrafast_2022}. Hence, we assume $I_\mathrm{c}(\mathbf{q})\approx0$. The $I_\mathrm{m}(\mathbf{q})$ were then corrected by dark images, normalized to the average FEL-pulse intensity and masked from parasitic scattering. We take the azimuthal average of $I_\mathrm{m}(\mathbf{q})$ which reduces the 2D~to a 1D~intensity distribution~(see Fig.\,\ref{Fig2_ScattAnalysis}\,\textbf{c}). By that, we treat the 2D labyrinth domain state as a 1D~chain of up- and down-magnetized domains with average domain characteristics. For analysis of the resulting magnetic scattering intensity $I_\mathrm{m}(q)$, we employ a Lorentzian empirical fitting function~\cite{zusin_ultrafast_2022}
\begin{equation}
I_\mathrm{m}(q) = \underbrace{e^{-2q/q_\mathrm{w}}}_{F(q)^2}\underbrace{\left[m_0 + \frac{m_1}{\left(\frac{q-q_1}{w_1} \right)^2 + 1} \right]^2}_{S(q)^2}.
\label{Eq_Lorentzian_Zusin}
\end{equation}
The first term outside of the square brackets is the form factor~$F(q)^2$ which is associated with the magnetic unit cell in real space. It is determined by the domain wall parameter $q_\mathrm{w}$ and accounts for the asymmetric shape of $I_\mathrm{m}(q)$. The term in the square brackets is the magnetic structure factor~$S(q)^2$ corresponding to the spatial arrangement of magnetic domains, i.\,e., the basic magnetic lattice in real space. It consists of a linear superposition of random uniform spatial fluctuations~$m_0$ and the first-order Lorentzian diffraction peak with amplitude~$m_1$, position~$q_1$ and linewidth~$w_1$. Let us emphasize that eq.\,(\ref{Eq_Lorentzian_Zusin}) is purely phenomenological. The same functionality, however, has been used and shown to fit scattering data from tr-XRMS up to the fifth diffraction order with excellent accuracy by substituting $S(q)^2$ with a sum of Lorentzian functions~\cite{zusin_ultrafast_2022}.

A fit of eq.\,(\ref{Eq_Lorentzian_Zusin}) to $I_\mathrm{m}(q, t=\SI{-1}{ps})$ is shown in Fig.\,\ref{Fig2_ScattAnalysis}\,\textbf{c} together with the individual contributions $F(q)^2$ and $S(q)^2$. An illustration of the individual contributions in real space is given in Fig.\,\ref{Fig2_ScattAnalysis}\,\textbf{d}. The exponential form factor contribution with $q_\mathrm{w}(t=\SI{-1}{ps})=0.1446\pm\SI{0.0118}{nm^{-1}}$ is interpreted as the domain wall width $\delta_\mathrm{m} = 2\pi q_\mathrm{w}^{-1} = 43.4\pm\SI{3.5}{nm}$. Labyrinth domain states in Co/Pt~multilayers with PMA exhibit strong Bloch domain wall character~\cite{riepp_multi-scale_2021} with a width defined by $\delta_\mathrm{B}=\pi\sqrt{A_\mathrm{ex}(|K_1 + K_2|)^{-1}}$~\cite{trauble_ferromagnetische_1965}. Using the measured $K_1=\SI{19.6}{kJ\,m^{-3}}$ and $K_2=\SI{-159.1}{kJ\,m^{-3}}$~(see Methods), as well as an exchange stiffness $A_\mathrm{ex}=\SI{23.3}{pJ\,m^{-1}}$ in Co/Pt~multilayers with PMA and an individual Co-layer thickness of $\SI{0.8}{nm}$~\cite{metaxas_creep_2007}, we obtain a calculated $\delta_\mathrm{B}=\SI{40.6}{nm}$ in good agreement with the $\delta_\mathrm{m}$ determined by XRMS. The magnetic structure of the labyrinth domain state is characterized by $q_1(t=\SI{-1}{ps})=0.0466\pm\SI{0.0004}{nm^{-1}}$ corresponding to an average domain period $\xi_\mathrm{m} = 2\pi/q_1 = 135.3 \pm \SI{1.2}{nm}$ and $w_1(t=\SI{-1}{ps})=0.0199\pm\SI{0.0011}{nm^{-1}}$ corresponding to a lateral correlation length $\lambda_\mathrm{m} = 2\pi/w_1 = 316.1 \pm \SI{17.5}{nm}$.
%

%
\begin{figure}
{\includegraphics[width=0.48\textwidth]{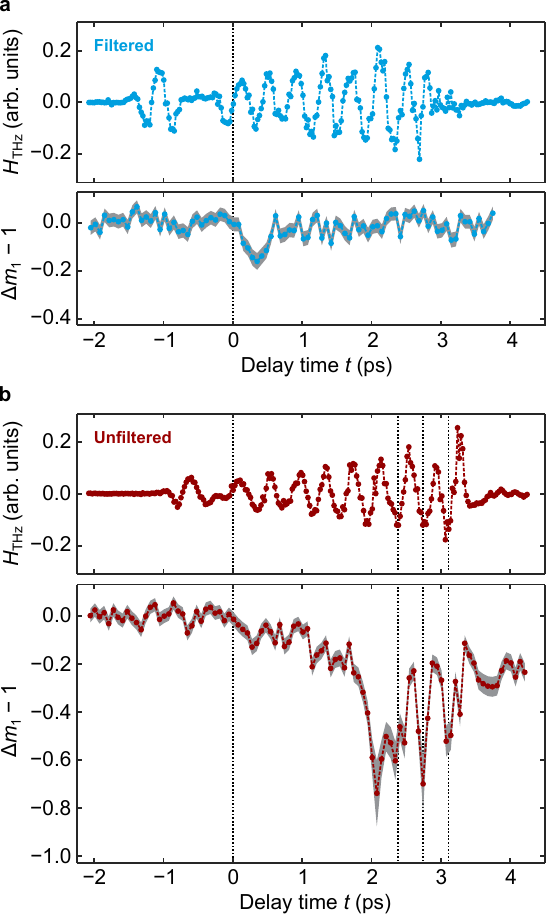}
\caption{\label{Fig3_MagnDyn} 
\textbf{THz-driven magnetization dynamics \quad a}~$H_\mathrm{THz}$-field trace determined from electro-optical sampling~(EOS) and transient $z$-component of the magnetization $\Delta m_1(t) - 1$ using the filtered THz-pump pulses (incident fluence $F_\mathrm{THz}<\SI{23}{mJ\,cm^{-2}}$). \textbf{b}~The same as \textbf{a} but for the unfiltered THz-pump pulses~(incident fluence $F_\mathrm{THz}\approx\SI{92}{mJ\,cm^{-2}}$). Gray-shaded areas are fit errors. Vertical dotted lines are guides to the eye.}}
\end{figure}
\textbf{THz-driven magnetization dynamics} In the following, we discuss the time evolution of the amplitude $m_1$ of the magnetic structure factor~$S(q)^2$ which corresponds to the $z$-component of the magnetization $m_z$. Relative changes $\Delta m_1(t) - 1$ are presented for the case of the filtered~($\nu \lesssim \SI{6.0}{THz}$) and the unfiltered THz~excitation in Fig.\,\ref{Fig3_MagnDyn}\,\textbf{a} and~\textbf{b}, respectively, together with the $H_\mathrm{THz}$-($E_\mathrm{THz}$-)field traces measured by electro-optical sampling~(EOS) before the respective measurements. Here, $\Delta m_1(t) = m_1(t)/\langle m_1(t<0)\rangle_t$.

The response of the $z$-component of the magnetization to the filtered THz-pump pulses is an ultrafast quenching by $16\%$ within $\tau_\mathrm{d} \approx \SI{400}{fs}$ followed by an equally fast and full recovery~(Fig.\,\ref{Fig3_MagnDyn}\,\textbf{a}). A maximum degree of demagnetization of $16\%$ agrees well with the observations in a \SI{15}{nm}~Co thin film pumped with a comparable fluence and can be explained by $E_\mathrm{THz}$-field driven ultrafast demagnetization~\cite{shalaby_coherent_2018}. According to time-dependent density functional theory~(TD-DFT), the $E_\mathrm{THz}$-field drives a coherent displacement of the electrons accompanied by a very efficient spin--orbit-coupling-(SOC-)mediated demagnetization~\cite{pellegrini_ab_2022}. Thereafter, one could expect a step-like reduction of $m_z$ with each half-cycle of the $E_\mathrm{THz}$~field, i.\,e., within $\tau_\mathrm{d} = 0.5/\nu_0 = \SI{200}{fs}$ per demagnetization step for monochromatic THz~radiation with $\nu_0 = \SI{2.5}{THz}$. We speculate that the differences in the ultrafast response originate from the polychromaticity of the THz~radiation ($\SI{0}{THz} < \nu \lesssim \SI{6.0}{THz}$) that causes a more incoherent demagnetization driven by the different $E_\mathrm{THz}$-field components. Employing the tranfer matrix method, we obtain an absorbed fluence of only about $\SI{0.7}{mJ\,cm^{-2}}$ for the highest frequency component $\nu = \SI{6.0}{THz}$. It is known that incoherent ultrafast demagnetization driven by low-fluence IR~laser pulses is governed by an efficient energy equilibration with the lattice on sub-picosecond time scales. Here, the efficient energy transfer among sub-systems could explain why no further demagnetization steps within the \SI{3.5}{ps}~pump-pulse duration but an ultrafast recovery is observed. Furthermore, the weak magnetic field of the low-fluence THz-pump pulses could explain the absence of $H_\mathrm{THz}$-field induced coherent oscillations of $m_z$. We show in the next paragraph that $\Delta m_1(t)-1$ can be modeled by the convolution of low-fluence incoherent ultrafast demagnetization and strongly damped coherent oscillations due to the presence of PMA.

The situation completely changes when exciting the Co/Pt~multilayer with the unfiltered THz-pump pulses~(Fig.\,\ref{Fig3_MagnDyn}\,\textbf{b}). Now, $m_z$ undergoes a 3-step demagnetization reaching a maximum degree of $75\%$ after $\SI{2}{ps}$. The recovery is governed by magnetization oscillations with an amplitude of about $\pm20\%$ in resonance with the THz~fundamental frequency $\nu_0 = \SI{2.5}{THz}$. The increase of the maximum degree of demagnetization can be explained by the additional frequency components $\nu > \SI{6.0}{THz}$ and the associated increase of the pump fluence. Employing the transfer matrix method as before, we obtain a 10~times higher absorbed fluence for $\nu = \SI{20.0}{THz}$ which is the highest frequency component with intensity $I_\mathrm{norm}(\nu)>0.01\cdot I_\mathrm{norm}(\nu_0)$. We note that the pump spectrum contains even higher-frequency components up to the IR~regime. The step-like demagnetization qualitatively agrees with the $E_\mathrm{THz}$-field driven coherent displacement of electrons accompanied by SOC-mediated demagnetization predicted by TD-DFT~\cite{pellegrini_ab_2022}. 
%
\begin{figure*}
{\includegraphics[width=0.98\textwidth]{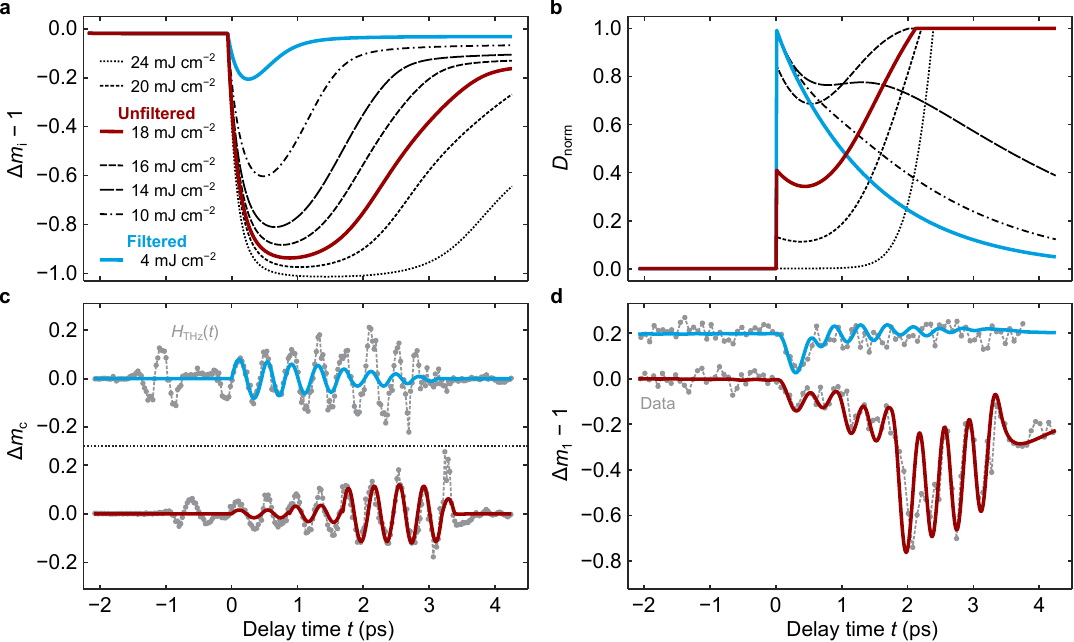}
\caption{\label{Fig4_ModelCohPrecess} 
\textbf{Phenomenological model \quad a}~Incoherent ultrafast demagnetization $\Delta m_\mathrm{i}(t) - 1$ determined from M3TM simulations using $F_\mathrm{i} = 4$--$\SI{24}{mJ\,cm^{-2}}$. Magnetization transients that were found to match the experimental data are shown in blue and red. \textbf{b}~Time-dependent damping $D_\mathrm{norm}(t)$ as given by eq.\,(\ref{eq_Damping}). Details in the text. \textbf{c}~Coherent magnetization oscillations $\Delta m_\mathrm{c}(t)=H_\mathrm{THz}(t)D_\mathrm{norm}(t)$. \textbf{d}~Final model for the transient $z$-component of the magnetization $\Delta m_1(t) - 1$ given by a convolution of the incoherent~(\textbf{a}) and coherent~(\textbf{c}) contributions.
}}
\end{figure*}
However, demagnetization steps with a duration of about $\SI{0.8}{ps} = 2/\nu_0$ are much longer than predicted, presumably due to a combination of demagnetization processes driven by the various $E_\mathrm{THz}$-field components. The high-frequency components thereby facilitate substantial energy transfer to the electron- and the spin-system reaching thermal equilibrium with the lattice on picosecond time scales. An onset of the oscillatory magnetization dynamics at $t=\SI{2}{ps}$ is rather surprising, as typically, an instantaneous response ($t=0$) is observed in THz-pump--probe experiments~(see, e.\,g., \cite{bonetti_thz-driven_2016, shalaby_coherent_2018, neeraj_terahertz_2022}). In comparison to these publications, we have to consider that the Co/Pt~multilayer exhibits PMA, i.\,e., an energetic minimum of aligning the magnetization along the $z$-direction. We show in the following paragraph that $\Delta m_1(t)-1$ can be modeled by the convolution of high-fluence incoherent ultrafast demagnetization and delayed coherent oscillations due to a heat-induced reduction of PMA.

\textbf{Phenomenological model} We model $\Delta m_1(t)$ as a convolution of incoherent ultrafast demagnetization $\Delta m_\mathrm{i}(t)$ and coherent magnetization oscillations $\Delta m_\mathrm{c}(t)$ consistent with, e.\,g.,~\cite{bonetti_thz-driven_2016, neeraj_terahertz_2022, liu_investigating_2022}
\begin{eqnarray}
\label{eq_ModelFun}
\Delta m_1(t) - 1 	&&= \left[\left( \Delta m_\mathrm{i}(t) - 1\right) \Theta(t) \right] \ast \Delta m_\mathrm{c}(t).
\end{eqnarray}
Here, $\Theta(t)$ is the Heaviside function accounting for the demagnetization onset at $t=0$. We treat the incoherent contribution $\Delta m_\mathrm{i}(t)$ as pure thermal demagnetization induced by an IR~pump pulse with $\lambda_\mathrm{i}=\SI{800}{nm}$. Obviously, this is an oversimplification as the filtered THz~spectrum does not contain any IR~components and the unfiltered THz~spectrum contains a broad frequency spectrum. In our approach we cast all $E_\mathrm{THz}$-field induced contributions, may they be coherent or incoherent electronic excitations, in one $\Delta m_\mathrm{i}(t)$ that is comparable to what is known from IR-induced ultrafast demagnetization. 

The incoherent contribution is simulated within the \textsf{udkm1Dsim} toolbox~\cite{schick_udkm1dsim_2021} that contains the microscopic three temperature model~(M3TM)~\cite{koopmans_explaining_2010} with heat diffusion along the sample $z$-direction~(see Methods). We simulated $\Delta m_\mathrm{i}(t)$ for a number of fluences and selected the transients that match the experimentally observed maximum degrees of demagnetization. This is the case for a fluence $F_\mathrm{i}=\SI{4}{mJ\,cm^{-2}}$ and $F_\mathrm{i} = \SI{18}{mJ\,cm^{-2}}$ when using the filtered and the unfiltered THz-pump pulses, respectively. The results from simulating $\Delta m_\mathrm{i}(t)$ via the M3TM are presented in Fig.\,\ref{Fig4_ModelCohPrecess}\,\textbf{a}. The electron- and phonon-temperature transients are provided in the extended data figures.

The coherent contribution $\Delta m_\mathrm{c}(t)$ is modeled via the product of the $H_\mathrm{THz}$-field trace and a time-dependent damping
\begin{equation}
\label{eq_Damping}
D(t) = e^{-\left( 1 - \frac{k_\mathrm{B}T_\mathrm{p}(t)}{K_1(t)V}\right) t},
\end{equation}
where $V=10\times10\times10\,\mathrm{nm^3}$ is the volume of a magnetic grain~(macrospin approximation). The phonon-temperature transients $T_\mathrm{p}(t)$ are known from the M3TM simulations and the anisotropy transients are calculated according to~\cite{bigot_ultrafast_2005}
\begin{equation}
\label{eq_Anisotropy}
K_1(T_\mathrm{p}(t)) =  K_1 m(T_\mathrm{p}(t))^{10}.
\end{equation}
We use $m(\tau)=[1-s\tau^{3/2}-(1-s)\tau^{5/2}]^{1/3}$, with the reduced temperature $\tau(t)=T_\mathrm{p}(t)/T_\mathrm{C}$, and $s=0.11$ for fcc Co~\cite{kuzmin_shape_2005}. The Curie temperature $T_\mathrm{C}=\SI{840}{K}$ was determined by vibrating sample magnetometry after the experiment~(see Methods). The calculated $K_1(T_\mathrm{p}(t))$ are provided in the extended data figures.

%
\begin{figure*}
{\includegraphics[width=0.98\textwidth]{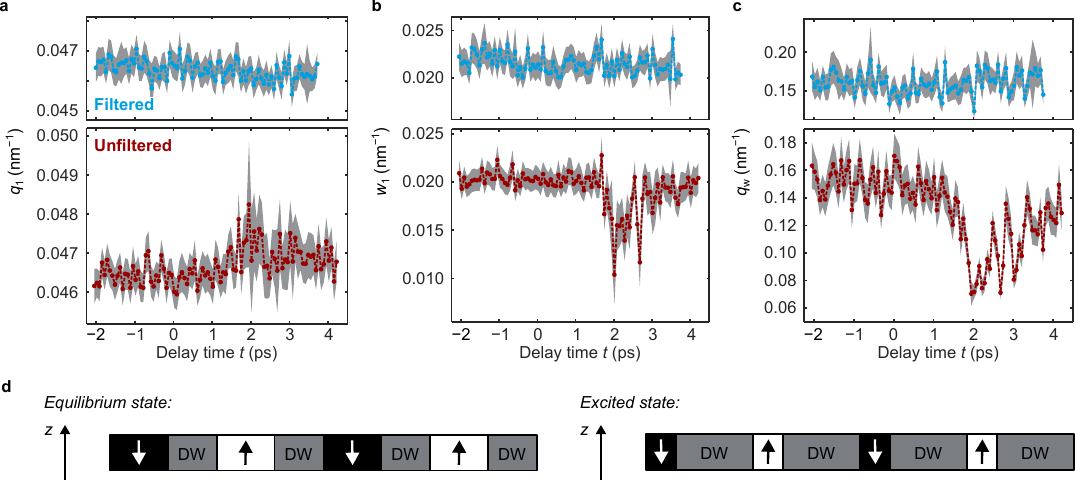}
\caption{\label{Fig5_DWDyn}
\textbf{THz-driven domain dynamics \quad a} and~\textbf{b}~Transient position $q_1(t)$ and width $w_1(t)$ of the domain state's structure factor. \textbf{c}~Transient domain-wall parameter $q_\mathrm{w}(t)$ of the domain state's form factor. Filtered and unfiltered scenarios are shown as blue and red data, respectively. Gray-shaded areas are fit errors. \textbf{d}~1D illustration of the equilibrium ($t<0$) and maximum excited domain state.
}}
\end{figure*}
In the limit of low fluences, $k_\mathrm{B}T_\mathrm{p}(t) \ll K_1(t)V$ at all times, i.\,e., the pump-induced heating of the lattice is too weak to induce a substantial reduction of PMA. In this case, $D(t)=D_\mathrm{norm}(t)$ becomes an exponential decay~(Fig.\,\ref{Fig4_ModelCohPrecess}\,\textbf{b}) and the coherent oscillations $\Delta m_\mathrm{c}(t)$ are strongly damped~(Fig.\,\ref{Fig4_ModelCohPrecess}\,\textbf{c}). In the limit of high fluences, $D(t)$ diverges, which corresponds to the unphysical case of strongly amplified oscillations. In case of $D(t)>1$, we therefore normalize eq.\,(\ref{eq_Damping}) to its value of minimum magnetic anisotropy $K_\mathrm{1,min}$ for $t < t(K_\mathrm{1,min})$ and set $D_\mathrm{norm}(t) = 1$ for $t > t(K_\mathrm{1,min})$. In other words, $D_\mathrm{norm}$ dynamically changes as $K_1$ decreases and reaches the regime of the undamped coherent oscillations ($D_\mathrm{norm}=1$) when $K_1 = K_\mathrm{1,min}$~(Fig.\,\ref{Fig4_ModelCohPrecess}\,\textbf{b}). The anisotropy $K_1$ decreases by about~75\% within the first~$\SI{2}{ps}$ while the coherent oscillations $\Delta m_\mathrm{c}(t)$ develop in amplitude~(Fig.\,\ref{Fig4_ModelCohPrecess}\,\textbf{c}). The convolutions of $\Delta m_\mathrm{i}(t)$ and $\Delta m_\mathrm{c}(t)$ are presented for the filtered and the unfiltered THz~radiation in Fig.\,\ref{Fig4_ModelCohPrecess}\,\textbf{d}. The model perfectly reproduces the features of both magnetization transients along the entire measured time range. For the unfiltered THz~radiation, larger deviations that exceed the experimental noise at $t\approx\SI{2}{ps}$ might be explained by the strong electromagnetic field that is predicted to lead to non-linearities in the magnetization response~\cite{hudl_nonlinear_2019}.

Note that for the case of a sample with negligible MAE, the criteria $D(t)>1$ holds from the start ($t=0$) and our model predicts an instantaneous (undamped) coherent response to the $H_\mathrm{THz}$~field as it was observed, e.\,g., in~\cite{bonetti_thz-driven_2016, shalaby_coherent_2018, neeraj_terahertz_2022}. Furthermore, it is consistent with the observation of an increasing delay of the coherent response with increasing MAE from fcc, bcc to hcp Co~\cite{unikandanunni_inertial_2022}. Even though our $D_\mathrm{norm}(t)$ is purely phenomenological, it is qualitatively in agreement with a time-dependent nutation damping factor derived from combining the time-dependent non-equilibrium Green function with the conventional Landau-Lifshitz-Gilbert~(LLG) formalism~\cite{bajpai_time-retarded_2019}. Their generalized LLG~equation contains a memory kernel that describes time retardation effects and originates from the fact that electron-spin can not follow instantaneously a change in the orientation of the local magnetic moments. It was even suggested in~\cite{unikandanunni_inertial_2022} that the coherent magnetization dynamics could be fully described by one time-dependent damping parameter that is qualitatively linked to a stronger electron--phonon scattering at sub-picosecond time scales and weaker spin--lattice relaxation at longer time scales.

\textbf{THz-driven domain dynamics} Finally, we investigate the effect of the THz-pump pulses on the lateral domain configuration, determined by the position $q_1$ and width $w_1$ of the structure factor as well as the domain-wall parameter $q_\mathrm{w}$ of the form factor~(Fig.\,\ref{Fig5_DWDyn}\,\textbf{a}--\textbf{c}). 

When using the filtered THz-pump pulses, constant fit parameters $q_1$, $w_1$ and $q_\mathrm{w}$ are obtained which is consistent with the fluence threshold for ultrafast domain dynamics observed when using IR~pump pulses~\cite{pfau_ultrafast_2012}. The parameters $q_1$, $w_1$ and $q_\mathrm{w}$ even remain constant for $t<\SI{2}{ps}$ when using high-fluence THz~pump pulses~(unfiltered) which demonstrates that both the domain structure and the domain walls maintain their equilibrium size-distribution on ultrafast time scales. This is qualitatively different to the ultrafast $q_1$-shift by $3$--$6\%$ to smaller values when using high-fluence IR-pump pulses~\cite{pfau_ultrafast_2012}. Originally explained by a broadening of the domain walls due to lateral superdiffusive spin transport, more recent experiments suggest ultrafast domain reconfigurations as an explanation, with a larger effect in low-symmetry systems like labyrinth domain states~\cite{zusin_ultrafast_2022, zhou_hagstrom_symmetry-dependent_2022}. However, no such ultrafast domain reconfigurations can be observed here, even for high-fluence THz~pump pulses. The absence of such ultrafast domain dynamics but rather the existence of a waiting time, that is determined by the time needed to compensate PMA, was reported for stripe domain states before~\cite{bergeard_irreversible_2015, lopez-flores_local_2020}. For a compensated PMA and in the presence of small IP~magnetic fields, the stripes were found to undergo a reorientation along the external field direction. Upon compensation of PMA after $t\approx\SI{2}{ps}$, here, the domain wall parameter $q_\mathrm{w}$ undergoes oscillatory dynamics that are highly correlated with the magnetization dynamics in Fig.\,\ref{Fig3_MagnDyn}\,\textbf{b}. Assuming that $q_\mathrm{w}$ inversely relates to the Bloch-wall width, this could be interpreted as a successive broadening and narrowing of the Bloch domain walls between \SI{43}{nm} and \SI{89}{nm} at maximum. A slight increase of $q_1$ within the error of the fit in combination with a sharp drop of $w_1$ to almost half its equilibrium value reveals an increased long-range order during these coherent oscillations from $\mathcal{O}=q_1/w_1\approx 2.3$ to $\mathcal{O}\approx 3.0$ at maximum. A situation where the domain-wall width increases while the average domain period remains largely the same is illustrated in Fig.\,\ref{Fig5_DWDyn}\,\textbf{d}. A high correlation between $m_1(t)$ and~$q_\mathrm{w}(t)$ for $t>\SI{2}{ps}$ is naturally convincing as, for oscillatory dynamics of the magnetization vector, a reduction of the $z$-component of the magnetization has to be associated with an increase of the $x$- and $y$-components and thus an increase of the domain-wall contribution in tr-XRMS. 
%

%
\section{Conclusions \label{IV}}
%
In conclusion, the magnetization of a Co/Pt~multilayer with PMA undergoes fluence-dependent dynamics upon excitation by polychromatic THz~pump pulses. These dynamics can be explained by a convolution of ultrafast demagnetization and coherent magnetization oscillations with time-dependent damping. For low pump fluences~(filtered), PMA causes a rapid alignment of $m_z$ along the $z$-direction, i.\,e., strongly damped coherent oscillations of $m_z$. For high pump fluences~(unfiltered), PMA undergoes a substantial reduction which enables undamped coherent oscillations of $m_z$ upon lattice heating. Our results demonstrate the existence of an upper speed limit for THz-driven magnetization switching in ferromagnets with PMA, i.\,e., a limit that is determined by the time needed to overcome the anisotropy energy barrier. It will be interesting to see if theoretical calculations can confirm a time-dependent nutation damping as the one proposed here. A reduction of the $m_z$ component during these coherent oscillations is associated with an increase in the $m_{x,y}$ components which, in tr-XRMS from a labyrinth domain state, is directly seen via highly correlated dynamics of the domain-wall parameter. The overall domain structure thereby remains largely unaffected, showing no signs of spin superdiffusion or ultrafast domain rearrangements, which highlights the applicability of THz~driven magnetization switching on the nanoscale. Our results thereby provide a guideline for controlling the THz-driven magnetization dynamics by tailoring PMA and changing the pump fluence. 

%
\section*{Acknowledgments}
%
We acknowledge DESY (Hamburg, Germany), a member of the Helmholtz Association HGF, for the provision of experimental facilities. Parts of this research were carried out at FLASH.  We thank S~D{\"u}sterer, M~Temme and the whole experimental team at FLASH for assistance in using the BL3~instrument. Beamtime was allocated for proposal F-20160531. We thank E~Jal, N~Bergeard and B~Vodungbo for many fruitful discussions as well as D~Hrabovsky at the MPBT platform of Sorbonne Université for his support with the VSM measurements. We acknowledge funding by the Deutsche Forschungsgemeinschaft (DFG) -- SFB-925 -- project ID~170620586, the Cluster of Excellence ‘Advanced Imaging of Matter’ of the DFG -- EXC-2056 -- project ID~390715994, the European Union’s Horizon 2020 research and innovation programme under the Marie Skłodowska-Curie grant agreement number 847471 and ANR-20-CE42-0012-01(MEDYNA).

%
\section*{Author contributions}
%
M.~R., A.~P.-K., L.~M., W.~R., R.~R., R.~F., K.~B., M.~W., R.~P., T.~G. and N.~S. performed the time-resolved experiments at FLASH and exploited the data. M.~R., A.~P.-K. and K.~B. grew the samples. M.~R., A.~P.-K., S.~M. and M.~H. performed the MOKE and VSM measurements. M.~R. conducted the simulations and wrote the paper. All authors discussed and improved the manuscript.
%
%
%
%
%

%
%
%
%
\appendix
\section*{Time-resolved XUV resonant magnetic scattering~(tr-XRMS)}
For the tr-XRMS experiment, FLASH was operated in the single-bunch mode providing \SI{60}{fs}~XUV~probe pulses at a repetition rate of \SI{10}{Hz}. The XUV~undulator was tuned to generate XUV~probe pulses with an average SASE spectrum centered around $\lambda_\mathrm{XUV}=\SI{20.8}{nm}$, i.\,e., a photon energy $E_\mathrm{XUV}=59.6\pm\SI{0.6}{eV}$ in resonance with the Co $M_{2,3}$~absorption edge~\cite{tiedtke_soft_2009}. Higher harmonics of the FEL~spectrum were blocked by a Si and Zr solid state filter which, in combination with the back-reflection focusing mirror, attenuate the probe-pulse intensity to about~$\SI{0.037}{\upmu J}$. With a beam size of~$52\times 40\,\mathrm{\upmu m^2}$, the calculated probe fluence is~$\SI{2.2}{mJ\,cm^{-2}}$. As expected for such a moderate fluence, no XUV-induced demagnetization nor XUV-induced permanent domain modifications were observed~\cite{kapcia_modeling_2022, riepp_-vacuum_2022}. A THz~beam with an about four times larger diameter than the XUV~beam ensured homogeneous excitation of the probed area. Diagnostic tools on the sample holder allowed for measuring coarse temporal as well as spatial overlap of the two beams at the sample position~\cite{riepp_multi-scale_2021}. The scattered intensity was recorded by a CCD with 2048$\times$2048~pixels and a pixel size of $\SI{13.5}{\upmu m}$. A beamstop-photodiode was installed centimeters from the detector to block the intense direct FEL beam and, at the same time, monitor FEL-intensity fluctuations for normalization of the data~\cite{muller_beam_2022}. The scattering statistics were improved by binning $4\times4$~pixels and accumulating 50~FEL~pulse exposures in one exposure of the CCD.

\section*{Sample Properties}
The sample used in this study was a ferromagnetic Pt(2.0)/[Co(0.8)/Pt(0.8)]$_8$/Pt(5.0)~multilayer grown on a Si$_3$N$_4$(50.0) multi-membrane substrate using sputtering techniques (numbers in nanometer). Structural investigations of Co/Pt~multilayers that were fabricated in the same way revealed polychrystallinity with pronounced (1\,1\,1)~texture and a grain size of about \SI{10}{nm}~\cite{winkler_variation_2015}. 

The first and second-order magnetic anisotropy constants $K_{1,2}$ were determined by magneto-optical Kerr effect~(MOKE) in polar and longitudinal geometry. Polar MOKE measurements revealed magnetic easy-axis behavior along the OOP~direction with a coercive field $\mu_0H_\mathrm{c}\approx\SI{25}{mT}$ and a saturation field $\mu_0H_\mathrm{s}\approx\SI{150}{mT}$. Longitudinal MOKE revealed magnetic hard-axis behavior along the IP~direction. $K_{1,2}$ were determined by fitting the (inverted) hard-axis hysteresis loop with
%
\begin{eqnarray}
\label{eq_MOKE}
\mu_0H_\mathrm{IP}(m_\parallel) = \frac{2K_1}{M_\mathrm{s}}m_\parallel + \frac{4K_2}{M_\mathrm{s}}m_\parallel^3,
\end{eqnarray}
where $M_\mathrm{s}=1.4\cdot10^{6}\,\mathrm{A\,m^{-1}}$ is the saturation magnetization in bulk Co at $T=\SI{0}{K}$ and $m_\parallel$ is the reduced magnetization component parallel to $H_\mathrm{IP}$. A fit of eq.\,(\ref{eq_MOKE}) to the data yields $K_1 = 19.6\pm4.7\,\mathrm{kJ\,m^{-3}}$ and $K_2 = -159.1\pm3.7\,\mathrm{kJ\,m^{-3}}$. The MOKE measurements and fit to the data are provided in the extended data figures. Prior to the FEL~beamtime, the sample was exposed to alternating OOP magnetic field cycles with decreasing amplitude and $\mu_0H_\mathrm{max}=\SI{1}{T}$ to generate a labyrinth domain state $m_z(\mathbf{r})$ close to the magnetic ground state. 

After the experiment, the temperature dependence of the saturation magnetization $M_\mathrm{s}(T)$ was measured employing vibrating sample magnetometry~(VSM) in an external magnetic field $\mu_0H_\mathrm{IP} = \SI{500}{mT}$. The temperature was increased from $T=\SI{300}{K}$ to $T=\SI{950}{K}$ at a rate $\Delta T = \SI{10}{K\,min^{-1}}$. The Curie temperature $T_\mathrm{C} \approx \SI{840}{K}$ was determined by a linear extrapolation of $M_\mathrm{s}(T)$ at high temperatures. The VSM measurement and the fit to the data are provided in the extended data figures.

%
%
%
\section*{M3TM Simulations}
%
%
\begin{table}[t]
\caption{\label{table_1}Material-specific parameters used for the M3TM simulations ($^*$\,assumtion)}
\begin{ruledtabular}
\begin{tabular}{cccc}
 & Co & Pt & Si$_3$N$_4$ \\ 
\hline \\
$C_\mathrm{e}\,(\mathrm{J\,kg^{-1}\,K^{-1}})$ & $0.0734~T_\mathrm{e}$~\cite{kuiper_spin-orbit_2014} & $0.0335~T_\mathrm{e}$~\cite{kuiper_spin-orbit_2014} & $0.0100~T_\mathrm{e}^*$ \\
$C_\mathrm{p}\,(\mathrm{J\,kg^{-1}\,K^{-1}})$ & $421$~\cite{haynes_crc_2014} & $133$~\cite{haynes_crc_2014} & $700$~\cite{haynes_crc_2014} \\
$\kappa_\mathrm{e}\,(\mathrm{W\,m^{-1}\,K^{-1}})$ & $20^*$ & $20^*$ & $20^*$ \\
$\kappa_\mathrm{p}\,(\mathrm{W\,m^{-1}\,K^{-1}})$ & $100$~\cite{haynes_crc_2014} & $71.6$~\cite{haynes_crc_2014} & $2.5$~\cite{ftouni_thermal_2015} \\
$\rho\,(\mathrm{kg\,m^{-3}})$ & 8860~\cite{haynes_crc_2014} & 21500~\cite{haynes_crc_2014} & 3190~\cite{haynes_crc_2014} \\
$n + ik$ & $2.53+4.88i$~\cite{werner_optical_2009} & $0.60+8.38i$~\cite{werner_optical_2009} & $2.00$~\cite{haynes_crc_2014}
\end{tabular}
\end{ruledtabular}
\end{table}
Incoherent ultrafast demagnetization is simulated within the \textsf{udkm1Dsim} toolbox~\cite{schick_udkm1dsim_2021} that contains the microscopic three temperature model~(M3TM) as proposed by B.~Koopmans \emph{et al.}~\cite{koopmans_explaining_2010}, including heat diffusion along the sample $z$-direction
%
\begin{eqnarray}
\label{eq_M3TM}
C_\mathrm{e}\rho\frac{\partial T_\mathrm{e}}{\partial t} &&= \frac{\partial}{\partial z}\left( \kappa_\mathrm{e}\frac{\partial T_\mathrm{e}}{\partial z}\right) - G_\mathrm{ep}\left( T_\mathrm{e} - T_\mathrm{p}\right) + S(z,t) \nonumber \\
C_\mathrm{p}\rho\frac{\partial T_\mathrm{p}}{\partial t} &&= \frac{\partial}{\partial z}\left( \kappa_\mathrm{p}\frac{\partial T_\mathrm{p}}{\partial z}\right) + G_\mathrm{ep}\left( T_\mathrm{e} - T_\mathrm{p}\right) \\
\frac{\partial m_\mathrm{i}}{\partial t} &&= Rm_\mathrm{i} \frac{T_\mathrm{p}}{T_\mathrm{C}}\left(1 - \coth\left(\frac{m_\mathrm{i}T_\mathrm{C}}{T_\mathrm{e}}\right)\right). \nonumber
\end{eqnarray}
The first two differentials describe the electron- and phonon-temperature transients, respectively, where $C_\mathrm{e}$ and $C_\mathrm{p}$ are the heat capacities, $\kappa_\mathrm{e}$ and $\kappa_\mathrm{p}$ are the thermal conductivities, $G_\mathrm{ep}$ is the electron--phonon coupling parameter and $\rho$ is the density. The initial heating of the electron system is given by the laser source term $S(z,t)$. Instead of a spin-temperature transient, the M3TM considers a magnetization transient that depends on $T_\mathrm{e}$ and $T_\mathrm{p}$, with a shape defined by $R=8a_\mathrm{sf}G_\mathrm{ep}k_\mathrm{B}T_\mathrm{C}^2V_\mathrm{at}\mu_\mathrm{B}\mu_\mathrm{at}^{-1}E_\mathrm{D}^{-2}$. Here, $a_\mathrm{sf}=0.15$ is the spin-flip probability, $k_\mathrm{B}$ is the Boltzmann constant, $T_\mathrm{C}=\SI{840}{K}$ is the Curie temperature, $V_\mathrm{at}=4\pi r_\mathrm{at}^3/3$ is the atomic volume with atomic radius $r_\mathrm{at}=\SI{1.35}{\AA}$, $\mu_\mathrm{at}/\mu_\mathrm{B}=1.72$ is the atomic magnetic moment in units of the Bohr magneton and $E_\mathrm{D}=\SI{0.0357}{eV}$ is the Debye energy of Co~\cite{koopmans_explaining_2010}. For the electron--phonon coupling parameter we take a constant value of $G_\mathrm{ep}=1.5\cdot10^{18}\,\mathrm{W\,m^{-3}\,K^{-1}}$ in Co~\cite{zahn_intrinsic_2022}. The \textsf{udkm1Dsim} toolbox yields a reflectivity of $85.6\%$ and a transmission of $4.5\%$ at $\lambda_\mathrm{i}=\SI{800}{nm}$, calculated by the transfer matrix method including multilayer absorption. 

Within the \textsf{udkm1Dsim} toolbox, in a first step, the Pt(2.0)/[Co(0.8)/Pt(0.8)]$_8$/Pt(6)/Si$_3$N$_4$(50) sample structure is generated as a 1D amorphous multilayer with material-specific properties for each subsystem~(see Table\,\ref{table_1}). In a second step, the laser source term $S(z,t)$ is defined as a delta-like pulse of high frequency ($\lambda_\mathrm{i}=\SI{800}{nm}$) with fluence $F_\mathrm{i} = 4$--$\SI{24}{mJ\,cm^{-2}}$. Note that the influence of the pump-pulse duration of $\SI{3.6}{ps}$ is taken into account via the coherent contribution $\Delta m_\mathrm{c}(t)$. In the final step, the \textsf{udkm1Dsim} toolbox calculates spatio-temporal heat-maps of the electron temperature, phonon temperature and magnetization for a certain delay range by solving eq.\,(\ref{eq_M3TM}) with an ODE solver. The $T_\mathrm{e}(t)$, $T_\mathrm{p}(t)$ and $\Delta m_\mathrm{i}(t)$ are obtained by taking the spatial average along the $z$-direction. The $T_\mathrm{e}(t)$, $T_\mathrm{p}(t)$ are provided in the extended data figures.
%

%
\section*{Extended Data Figures}
%
\begin{figure*}[h]
{\includegraphics[width=0.98\textwidth]{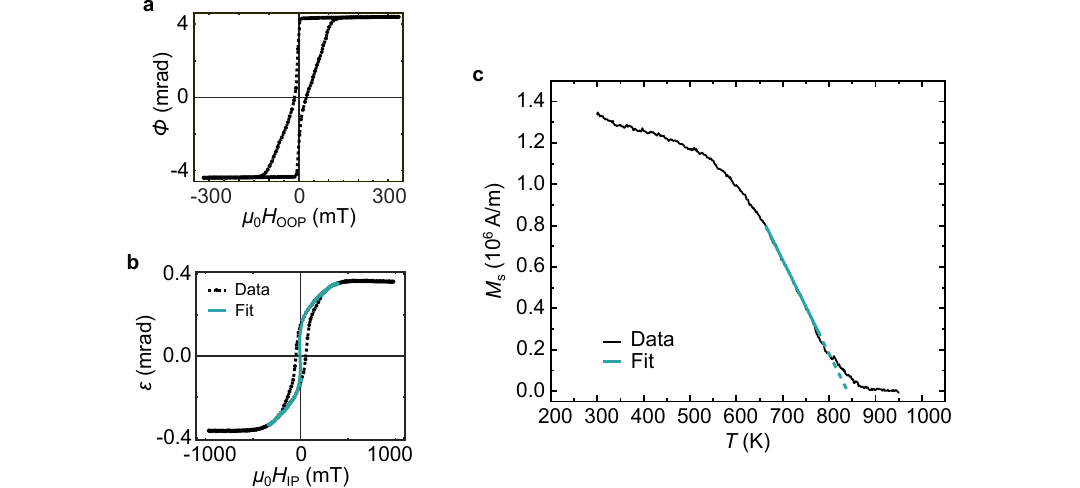}}
\caption{\label{Fig6_SampleProp} 
\textbf{Static magnetic properties of the [Co/Pt]$_8$~multilayer \quad a}~Polar and \textbf{b}~longitudinal MOKE at room temperature. The solid line in \textbf{b}~is a fit to the inverted data $\mu_0H_\mathrm{IP}(\varepsilon)$ for small $\varepsilon$ (details in the main article). \textbf{c}~Temperature dependence of the sponatneous magnetization measured by VSM in external magnetic field $\mu_0H_\mathrm{IP} = \SI{500}{mT}$. The Curie temperature $T_\mathrm{C} \approx \SI{840}{K}$ is determined by a linear extrapolation at high temperatures.}
\end{figure*}
%

%
\begin{figure*}[h]
{\includegraphics[width=0.98\textwidth]{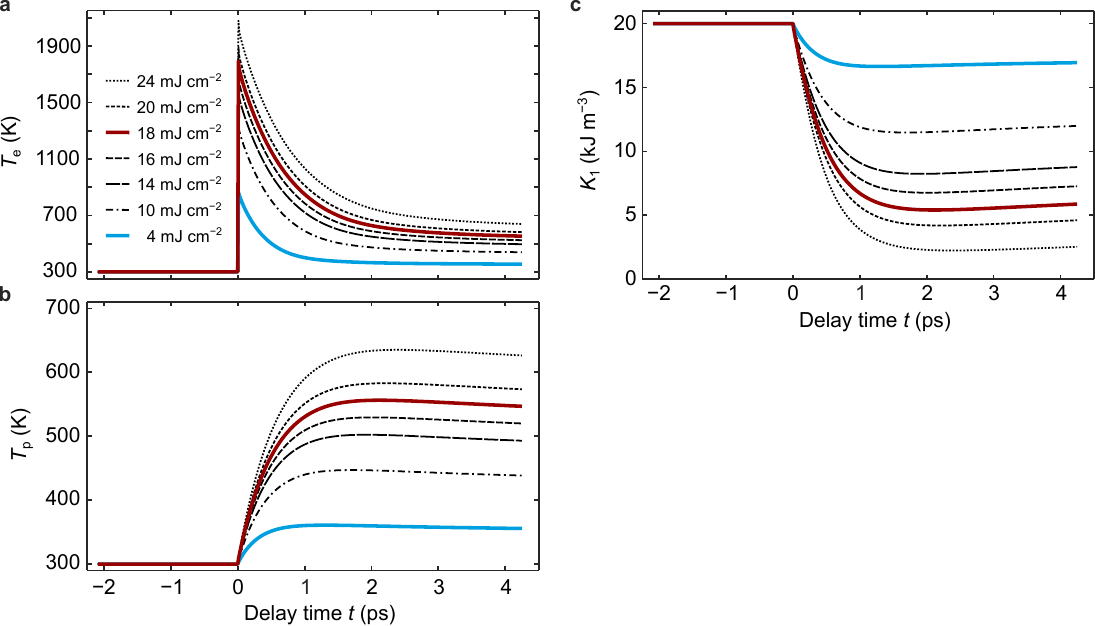}
\caption{\label{Fig7_M3TM} 
\textbf{Results from M3TM simulations of the [Co/Pt]$_8$ multilayer \quad a}~Electron-temperature transient $T_\mathrm{e}(t)$ and \textbf{b}~phonon-temperature transient $T_\mathrm{p}(t)$ for fluences $F_\mathrm{i} = 4$--$\SI{24}{mJ\,cm^{-2}}$. The transients are extracted from spatio-temporal heat maps averaged along the sample $z$-direction using the \textsf{udkm1Dsim} toolbox. \textbf{c}~First-order magnetic anisotroy transient $K_1(t)$ calculated as described in the main article.}}
\end{figure*}
\end{document}